\documentstyle[12pt]{article}
\def\Re{{\cal R \mskip-4mu \lower.1ex \hbox{\it e}\,}}
\def\Im{{\cal I \mskip-5mu \lower.1ex \hbox{\it m}\,}}
\def\ie{{\it i.e.}}
\def\eg{{\it e.g.}}

\def\etal{{\it et al.}}

\def\beq{\begin{equation}}
\def\eeq{\end{equation}}
\def\brbsg{Br(b\to s\gamma)}
\def\bsg{b\to s\gamma}
\def\Zbb{Z\rightarrow b\ov b}

\def\eps1{\epsilon_1}
\def\eps2{\epsilon_2}
\def\eps3{\epsilon_3}
\def\sub#1{_{\lower.25ex\hbox{$\scriptstyle#1$}}}
\def\sul#1{_{\kern-.1em#1}}
\def\sll#1{_{\kern-.2em#1}}
\def\sbl#1{_{\kern-.1em\lower.25ex\hbox{$\scriptstyle#1$}}}
\def\ssb#1{_{\lower.25ex\hbox{$\scriptscriptstyle#1$}}}
\def\sbb#1{_{\lower.4ex\hbox{$\scriptstyle#1$}}}

\def\tev{\,{\rm TeV}}
\def\GeV{\,{\rm GeV}}

\def\JL{J. L. Lopez}
\def\DVN{D. V. Nanopoulos}

\def\to{\rightarrow}
\def\ov{\overline}
\def\mh{\ifmmode m\sbl H \else $m\sbl H$\fi}
\def\mch{\ifmmode m_{H^\pm} \else $m_{H^\pm}$\fi}
\def\mt{\ifmmode m_t\else $m_t$\fi}
\def\mc{\ifmmode m_c\else $m_c$\fi}
\def\mz{\ifmmode M_Z\else $M_Z$\fi}
\def\mw{\ifmmode M_W\else $M_W$\fi}
\def\mws{\ifmmode M_W^2 \else $M_W^2$\fi}
\def\mhs{\ifmmode m_H^2 \else $m_H^2$\fi}
\def\mzs{\ifmmode M_Z^2 \else $M_Z^2$\fi}
\def\mts{\ifmmode m_t^2 \else $m_t^2$\fi}
\def\mcs{\ifmmode m_c^2 \else $m_c^2$\fi}
\def\mchs{\ifmmode m_{H^\pm}^2 \else $m_{H^\pm}^2$\fi}
\def\ztwo{\ifmmode Z_2\else $Z_2$\fi}
\def\zone{\ifmmode Z_1\else $Z_1$\fi}
\def\mtwo{\ifmmode M_2\else $M_2$\fi}
\def\mone{\ifmmode M_1\else $M_1$\fi}
\def\tb{\ifmmode \tan\beta \else $\tan\beta$\fi}
\def\xw{\ifmmode x\sub w\else $x\sub w$\fi}
\def\ch{\ifmmode H^\pm \else $H^\pm$\fi}
\def\lum{\ifmmode {\cal L}\else ${\cal L}$\fi}
\def\inpb{\ifmmode {\rm pb}^{-1}\else ${\rm pb}^{-1}$\fi}
\def\infb{\ifmmode {\rm fb}^{-1}\else ${\rm fb}^{-1}$\fi}
\def\epem{\ifmmode e^+e^-\else $e^+e^-$\fi}
\def\ppb{\ifmmode \bar pp\else $\bar pp$\fi}

\newskip\zatskip \zatskip=0pt plus0pt minus0pt

\def\matth{\mathsurround=0pt}
\def\lsim{\mathrel{\mathpalette\atversim<}}
\def\gsim{\mathrel{\mathpalette\atversim>}}
\def\atversim#1#2{\lower0.7ex\vbox{\baselineskip\zatskip\lineskip\zatskip
  \lineskiplimit 0pt\ialign{$\matth#1\hfil##\hfil$\crcr#2\crcr\sim\crcr}}}

\renewcommand{\thefootnote}{\fnsymbol{footnote}}

\begin{document} \begin{titlepage}
\setcounter{page}{1}
\thispagestyle{empty}
\rightline{\vbox{\halign{&#\hfil\cr
&YUMS-96-2\cr
&SNUTP-96-003\cr
&hep-ph/9607343\cr
&July 1996\cr}}}
\vspace{0.1in}
\begin{center}
\vglue 0.3cm
{\Large\bf $b\rightarrow s\gamma$ and $\epsilon_1$ Constraints on\\}
\vspace{0.2cm}
{\Large\bf Supergravity Models\\}
\vglue 1.5cm
{GYE T. PARK\\}
\vglue 0.4cm
{\em Department of Physics, Yonsei University\\}
{\em Seoul, 120-749, Korea\\}
\baselineskip=12pt

\end{center}

\begin{abstract}

In the light of the top quark discovered very recently by CDF,
we investigate the possibility of narrowing down
the allowed top quark masses by combining 
for the first time only two
strongest constraints present in the no-scale $SU(5)\times U(1)$ supergravity model, namely, the ones from the flavor-changing radiative decay $b\rightarrow s\gamma$ and the precision
measurements at LEP in the form of $\epsilon_{1}$.
It turns out that even without including the most devastating constraint from
$\Zbb$ measurement at LEP in the form of $R_b$ directly or $\epsilon_b$
indirectly, the combined constraint from $b\rightarrow s\gamma$ and $\epsilon_1$ alone in fact excludes $m_t(m_t)\gsim 180\GeV$
altogether in the no-scale model, providing a constraint on $m_t$
near the upper end of the CDF values.
The resulting upper bound on $m_t$ is stronger and $5 \GeV$ lower than the one 
from combining $\epsilon_1$ and $\epsilon_b$ constraints and also combining
$b\rightarrow s\gamma$ and $\epsilon_b $ constraints in the previous analysis.

\end{abstract}

\renewcommand{\thefootnote}{\arabic{footnote}} \end{titlepage}
\setcounter{page}{1}


With the top quark discovered very recently by the CDF Collaboration from Fermi Laboratory
in $\ov{p} p$ collisions
with the measured top mass \cite{CDF-topdiscovery}, $m_t=176\pm 8\pm10\;{\rm GeV}$, the Standard Higgs mass $m_H$ is now 
the only unknown parameter in the Standard Model(SM).
Despite the remarkable successes of the SM in its complete
agreement with current experimental data, there is still no
experimental information on the nature of its Higgs sector.
The unknown $m_t$ has long been one of the biggest obstacles in studying the phenomenology of the SM and its extensions of interest.
Now that $m_t$ is measured, one should be able to narrow down
the values of $m_t$ to the vicinity of the above central value
considering the large experimental uncertainties in the measured top mass.
In the context of supersymmetry, such a task can be performed  within the Minimal Supersymmetric Standard Model (MSSM)
\cite{OldEW,BFC,ABC}. The problem with such calculations is that there are too many parameters in the MSSM and therefore it is not possible to obtain precise predictions for the observables of interest.
In the context of supergravity models (SUGRA), on the other hand, any observable can be
computed in terms of at most five parameters: the top-quark mass, the ratio of
Higgs vacuum expectation values ($\tan\beta$), and three universal
soft-supersymmetry-breaking parameters $(m_{1/2},m_0,A)$. This
implies much sharper predictions for the various quantities of interest, as
well as numerous correlations among them. Of even more experimental interest is
$SU(5)\times U(1)$ SUGRA where string-inspired ans\"atze for the soft-supersymmetry-breaking parameters allow the theory to be described in terms of only three parameters: $m_t$, $\tan\beta$, and $m_{\tilde g}$.

In this letter, we would like to present the possibility of narrowing down
the allowed top quark masses by combining for the first time
two strongest constraints present in the no-scale $SU(5)\times U(1)$ SUGRA, namely, the ones from the flavor-changing radiative decay $b\rightarrow s\gamma$ and the precision
measurements at LEP in the form of $\epsilon_{1}$.
After the first observation by CLEO on the exclusive decay $B\rightarrow K^*\gamma$ \cite{Thorndike} which provides only the upper bound on
the inclusive branching ratio of $b\rightarrow s\gamma$ decay
, CLEO has recently measured 
the inclusive branching ratio of $b\rightarrow s\gamma$ decay \cite{CLEO94}
, which can provide much more precise contraint.
Among four variables
$\epsilon_{1,2,3,b}$ introduced in Ref.~\cite{BFC,ABC}, only $\epsilon_1$ and $\epsilon_b$ lead to
significant constraints in supersymmetric models \cite{{eps-sugra,kimparkRb}}.
Although $\epsilon_b$, which encodes the vertex corrections to
$\Zbb$, provides even stronger contraint than $\epsilon_1$
for $m_t< 170\GeV$ \cite{kimparkRb,ProbingKP},
we do not include in our analysis here the $\epsilon_b$ constraint
in an attempt to isolate and exclude the most devastating impact 
on the precision test
from the recent LEP measurement on $R_b$ ($\equiv{\Gamma(Z\rightarrow b\ov b)\over{\Gamma(Z\rightarrow hadrons)}}$) whose latest experimental values \cite{RbLP95} lie more than three standard deviations above the SM predictions
for the values of $m_t$ from the CDF. In fact, the possibility of improving the situation to a certain extent
in this so called ``$R_b$- crisis'' has been studied in the context of SUGRA in Ref.~\cite{kimparkRb} \footnote{ As can be seen in  Ref.~\cite{kimparkRb} by imposing the latest experimental data on $R_b$ alone,
$m_t(m_t)\gsim 170\GeV$ may be excluded at $99$\% C.~L. in the no-scale
$SU(5)\times U(1)$ SUGRA.}.
An analysis similar to the one here has been done previously in Ref.~\cite{ProbingKP} where the combined constraints 
from $\epsilon_1$ and $\epsilon_b$  and also from
$b\rightarrow s\gamma$ and $\epsilon_b $ were studied.
However, the correlated constraint from $b\rightarrow s\gamma$ and $\epsilon_1$
has never been studied.

The $SU(5)\times U(1)$ SUGRA contains, at low energy, the
SM gauge symmetry and the particle content of the MSSM. The procedure to restrict 5-dimensional parameter spaces is as follows \cite{aspects}.
First, upon sampling a specific choice of ($m_{1/2},m_0,A$) at the unification scale and ($m_t,\tan\beta$) at the electroweak scale, the renormalization group equations (RGE) are run from the unification scale to the electroweak scale, where the radiative electroweak breaking condition is imposed by minimizing the effective 1-loop Higgs potential, which determines the Higgs mixing term $\mu$ up to its sign.
Here the sign of $\mu$ is given as usual \cite{signmu}, and differs from that of Ref.~\cite{kimparkRb,bsgamma}; {\it i.e.}    , we define $\mu$ by $W_\mu =\mu H_1 H_2$.

We also impose consistency constraints such as perturbative unification and the naturalness bound of $m_{\tilde g}\lsim 1\tev$.
Finally, all the known experimental bounds on the sparticle masses are imposed\footnote{We use the following experimental lower bounds on the sparticle masses in $\GeV$ in the order of gluino, squarks, lighter stop, sleptons, and
lighter chargino: $m_{\tilde g}\gsim 150$, $m_{\tilde q}\gsim 100$,
$m_{{\tilde{t}}_1}\gsim 45$, $m_{\tilde l}\gsim 43$, 
$m_{\chi^\pm_1}\gsim 45$.}.  
This prodedure yields the restricted parameter spaces for the model.
Further reduction in the number of input parameters in $SU(5)\times U(1)$ SUGRA is made possible because in specific string-inpired scenarions for ($m_{1/2},m_0,A$) at the unification scale these three parameters are computed in terms of just one of them \cite{IL}. One obtains $m_0=A=0$ in the {\em no-scale} scenario.

The inclusive branching ratio of $b\rightarrow s\gamma$ decay
has been recently measured for the first time by CLEO 
to be at 95\% C.~L. \cite{CLEO94},
$$1\times 10^{-4}<Br(b\rightarrow s\gamma) <4\times 10^{-4}.$$
This follows the renewed surge of interests on the $\bsg$ decay, spurred by the
CLEO bound $\brbsg<8.4\times10^{-4}$ at $90\%$ C.L. \cite{CLEO}, with which
it was pointed out in Ref.~ \cite{BargerH} that the CLEO bound can be violated due to the charged Higgs
contribution in the 2 Higgs doublet  model (2HDM) and the MSSM basically if $m_{H^\pm}$ is too light, excluding large portion of the charged Higgs parameter space. It has certainly proven that this particular decay mode can provide more stringent constraint on new physics beyond SM than any other experiments\cite{ProbingKP,bsgamma,Rbbsg2HD}. As we know, with the increasing accuracy of the LEP measurements, it has become extremely important performing the precision test of the SM and its extensions\footnote{A standard model fit to the latest LEP data yields the top mass,
$m_t=178\pm 11^{+18}_{-19}\;{\rm GeV}    $  \cite{Schaile}, which is in perfect agreement with the measured top mass from CDF.}.
Among several different schemes to analyze precision electroweak tests, we choose a scheme introduced by Altarelli \etal
\cite{ABC,Altlecture} where four variables, $\epsilon_{1,2,3}$ and $\epsilon_b$
are defined in a model independent way. These four variables correspond to 
a set of observables $\Gamma_{l}, \Gamma_{b}, A^{l}_{FB}$ and $M_W/M_Z$.
Among these variables, $\epsilon_b$ encodes the vertex corrections to
$Z\rightarrow b\overline b$.

In the 2HDM and the MSSM, $b\rightarrow s\gamma    $ decay receives significant contributions from penguin diagrams with $W^\pm-t$ loop, $H^\pm-t$ loop \cite{HCbsg} and the $\chi^\pm_{1,2}-\tilde t_{1,2}$ loop \cite{Bertolini}
only in the MSSM.
The expression used for $Br(b\rightarrow     s\gamma)    $ in the leading logarithmic (LL) calculations is given by \cite{GSW}
\begin{equation}    
{Br(b\rightarrow     s\gamma)\over Br(b\rightarrow     ce\bar\nu)}={6\alpha\over\pi}
{\left[\eta^{16/23}A_\gamma
+{8\over3}(\eta^{14/23}-\eta^{16/23})A_g+C\right]^2\over
I(m_c/m_b)\left[1-{2\over{3\pi}}\alpha_s(m_b)f(m_c/m_b)\right]},\label{bsg}
\end{equation}    
where $\eta=\alpha_s(M_W)/\alpha_s(m_b)$, $I$ is the phase-space factor
$I(x)=1-8x^2+8x^6-x^8-24x^4\ln x$, and $f(m_c/m_b)=2.41$ the QCD
correction factor for the semileptonic decay.
$C$ represents the leading-order QCD
corrections to the $b\rightarrow s\gamma    $ amplitude when evaluated at the $\mu=m_b$ scale
\cite{GSW}.
We use the 3-loop expressions for $\alpha_s$ and choose $\Lambda_{QCD}$ to
obtain $\alpha_s(M_Z)$ consistent with the recent measurements at LEP.
In our computations we have used: $\alpha_s(M_Z)=0.118$, $ Br(b\rightarrow     ce\bar\nu)=10.7\%$, $m_b=4.8\;{\rm GeV}    $, and
$m_c/m_b=0.3$. The $A_\gamma,A_g$ are the
coefficients of the effective $bs\gamma$ and $bsg$ penguin operators
evaluated at the scale $M_W$. Their simplified expressions are given in Ref.~\cite{BG}
in the justifiable limit of negligible gluino and neutralino contributions \cite{Bertolini} and degenerate squarks, except for the $\tilde t_{1,2}$ which are significantly split by $m_t$.
Regarding large uncertainties in the LL QCD corrections, which is mainly due to the choice of renormalization scale $\mu$ and is estimated to be $\approx 25\%$, it has been recently demonstrated by Buras {\it et al.} in Ref.~\cite{burasetal}
that the significant $\mu$ dependence in the LL result can in fact be reduced
considerably by including next-to-leading logarithmic (NLL) corrections, which however, involves very complicated calculations of three-loop mixings
between cetain effective operators and therefore have not been completed yet.

The expression for $\epsilon_1$ is given as
\cite{BFC}
\beq
\epsilon_1=e_1-e_5-{\delta G_{V,B}\over G}-4\delta g_A,\label{eps1}
\eeq
where $e_{1,5}$ are the following combinations of vacuum polarization
amplitudes
\begin{eqnarray}
e_1&=&{\alpha\over 4\pi \sin^2\theta_W M^2_W}[\Pi^{33}_T(0)-\Pi^{11}_T(0)],
\label{e1}\\
e_5&=& M_Z^2F^\prime_{ZZ}(M_Z^2),\label{e5}
\end{eqnarray}
and the $q^2\not=0$ contributions $F_{ij}(q^2)$ are defined by
\beq
\Pi^{ij}_T(q^2)=\Pi^{ij}_T(0)+q^2F_{ij}(q^2).
\eeq
The $\delta g_A$ in Eq.~(\ref{eps1}) is the contribution to the axial-vector
form factor at $q^2=M^2_Z$ in the $Z\to l^+l^-$ vertex from proper vertex
diagrams and fermion self-energies, and $\delta G_{V,B}$ comes from the
one-loop box, vertex and fermion self-energy corrections to the $\mu$-decay
amplitude at zero external momentum. These non-oblique SM corrections are
non-negligible, and must be included in order to obtain an accurate SM
prediction.
As is well known, the SM contribution to $\epsilon_1$ depends quadratically
on $m_t$ but only logarithmically on the SM Higgs boson mass ($m_H$). In this
fashion upper bounds on $m_t$ can be obtained which have a non-negligible $m_H$
dependence: up to $20\GeV$ stronger when going from a heavy ($\approx 1\tev$)
to a light ($\approx100\GeV$) Higgs boson. It is also known (in the MSSM) that
the largest supersymmetric contributions to $\epsilon_1$ are expected to
arise from the $\tilde t$-$\tilde b$ sector, and in the limiting case of a very
light stop, the contribution is comparable to that of the $t$-$b$ sector. The
remaining squark, slepton, chargino, neutralino, and Higgs sectors all
typically contribute considerably less. For increasing sparticle masses, the
heavy sector of the theory decouples, and only SM effects  with a {\it light}
Higgs boson survive. (This entails stricter upper bounds on $m_t$ than in the
SM, since there the Higgs boson does not need to be light.) However, for a
light chargino ($m_{\chi^\pm_1}\to{1\over2}M_Z$), a $Z$-wavefunction
renormalization threshold effect can introduce a substantial $q^2$-dependence
in the calculation, \ie, the presence of $e_5$ in Eq.~(\ref{eps1}) \cite{BFC}.
The complete vacuum polarization contributions from the Higgs sector, the
supersymmetric chargino-neutralino and sfermion sectors, and also the
corresponding contributions in the SM have been included in our calculations.

In Fig.~1 we show the results of the calculation of $\epsilon_1$
for all allowed points in the no-scale $SU(5)\times U(1)$ SUGRA for the running top mass $m_t(m_t)= 170, 180\GeV$.
We use in the figure the following experimental value for $\epsilon_1$,
$$\epsilon_1^{exp}=(3.5\pm 1.8)\times 10^{-3},$$ determined from the latest $\epsilon$- analysis using the LEP and SLC data in Ref.~\cite{Altarelli94}. 
In the figure points between the two horizontal lines are allowed by the $\epsilon_1$ constraint at the 90\% C.~L. 
Since all sparticle masses nearly scale with the gluino mass (or the chargino mass), it suffices to show the dependences of the parameter on, for example, the chargino mass.
Therefore, we show the explicit dependence only on the chargino mass in Fig.~1. The significant drop in $\epsilon_1$ comes from the threshold effect of Z-wavefunction renormalization as discussed earlier.
As can be seen in Fig.~1, the current experimental values for $\epsilon_{1}$ prefer light but not too light chargino in the no-scale model: 
$$50\GeV\lsim m_{\chi^\pm_1}\lsim 70 \GeV, \qquad m_t(m_t)=180\GeV ,$$
$$50\GeV\lsim m_{\chi^\pm_1}, \qquad m_t(m_t)=170\GeV ,$$
In order to deduce the bounds on any of the other masses from the above bounds, one can use the scaling relations in the model,
$m_{\tilde q}\approx0.97m_{\tilde g}$ and
$2m_{\chi^0_1}\approx m_{\chi^0_2}\approx m_{\chi^\pm_1}\approx
0.28 m_{\tilde g}$.
For $m_t(m_t)< 170\GeV$, it is fairly clear that there will be only lower bound
on the chargino mass.
In addition to the $\epsilon_1$ constraint there is another very strong constraint coming from the $\bsg$ decay as mentioned earlier.
In order to see if it provides additional contraint, in Fig.~2 we show
the results of the calculation of $\brbsg$ versus $\epsilon_1$.
It is very interesting for one to see that the combined constraint
is much stronger than each individual constraint and it
nearly excludes $m_t(m_t)= 180\GeV$ in the model, leaving only one point (out of a few thousand points) still allowed for $\mu<0$. 
However, this one point is also excluded by imposing
the preliminary but conservative bound from all 4 LEP collaborations
on the lightest chargino mass,
$m_{\chi^\pm_1}\gsim 65 \GeV$ \cite{LEP1.5}.
Therefore, $m_t(m_t)\gsim 180\GeV$ is excluded altogether in the no-scale model, and  
this provides a constraint on $m_t$
near the upper end of the CDF values.
Without imposing the new bound $m_{\chi^\pm_1}\gsim 65 \GeV$,
the resulting constraint is in fact even stronger than the one from
the combined constraints 
from $\epsilon_1$ and $\epsilon_b$  and also from
$b\rightarrow s\gamma$ and $\epsilon_b$ in Ref.~\cite{ProbingKP}.
On the other hand, the combined constraint for $m_t(m_t)= 170 \GeV$ becomes less severe
although the $\bsg$ constraint excludes additionally a large fraction of the parameter space.
The large suppression in $Br(b\rightarrow     s\gamma)    $ for $\mu<0$ in the model is worth
further explanation. As first noticed in Ref. \cite{bsgamma},
what happens is that in Eq.~(\ref{bsg}), the $A_\gamma$ term nearly
cancels against the QCD correction factor $C$; the $A_g$ contribution is small.
The $A_\gamma$ amplitude receives three contributions: from the $W^\pm$-$t$ loop, from the $H^\pm$-$t$ loop, and from the
$\chi^\pm_{1,2}$-${\tilde t}_{1,2}$ loop. The first two contributions are
always negative \cite{GSW}, whereas the last one can have either sign, making it possible having cancellations among three contributions.

In conclusion, in the light of the top quark discovered very recently by CDF,
we investigate the possibility of narrowing down
the allowed top quark masses by combining 
for the first time only two
strongest constraints present in the no-scale $SU(5)\times U(1)$ supergravity model, namely, the ones from the flavor-changing radiative decay $b\rightarrow s\gamma$ and the precision
measurements at LEP in the form of $\epsilon_{1}$.
It turns out that even without including the most devastating constraint from
$\Zbb$ measurement at LEP in the form of $R_b$ directly or $\epsilon_b$
indirectly, the combined constraint from $b\rightarrow s\gamma$ and $\epsilon_1$ alone in fact excludes $m_t(m_t)\gsim 180\GeV$
altogether in the no-scale model, providing a constraint on $m_t$
near the upper end of the CDF values.
The resulting upper bound on $m_t$ is stronger and $5 \GeV$ lower than the one 
from combining $\epsilon_1$ and $\epsilon_b$ constraints and also combining
$b\rightarrow s\gamma$ and $\epsilon_b $ constraints in the previous analysis.

\vskip.25in
\centerline{ACKNOWLEDGEMENTS}

This work has been supported in part by
NON DIRECTED RESEARCH FUND, Korea Research Foundation, in part by Yonsei University Faculty Research Grant, and in part by Center for Theoretical Physics, Seoul National University. 


%
\def\NPB#1#2#3{Nucl. Phys. B {\bf#1} (19#2) #3}
\def\PLB#1#2#3{Phys. Lett. B {\bf#1} (19#2) #3}
\def\PLIBID#1#2#3{B {\bf#1} (19#2) #3}
\def\PRD#1#2#3{Phys. Rev. D {\bf#1} (19#2) #3}
\def\PRL#1#2#3{Phys. Rev. Lett. {\bf#1} (19#2) #3}
\def\PRT#1#2#3{Phys. Rep. {\bf#1} (19#2) #3}
\def\MODA#1#2#3{Mod. Phys. Lett. A {\bf#1} (19#2) #3}
\def\IJMP#1#2#3{Int. J. Mod. Phys. A {\bf#1} (19#2) #3}
\def\TAMU#1{Texas A \& M University preprint CTP-TAMU-#1}
\def\ARAA#1#2#3{Ann. Rev. Astron. Astrophys. {\bf#1} (19#2) #3}
\def\ARNP#1#2#3{Ann. Rev. Nucl. Part. Sci. {\bf#1} (19#2) #3}

\newpage

%
{\bf Figure Captions}
\begin{itemize}

\item Figure 1: The predictions for $\epsilon_1$ versus the chargino mass in the no-scale $SU(5)\times
U(1)$ SUGRA for the running top mass 
$m_t=180\GeV$ (top row) and $m_t=170\GeV$ (bottom row). In the figure, points
between the horizontal lines are allowed at the 90\% C.~L.  
\item Figure 2: The correlated predictions for $BR(\equiv Br(b\rightarrow     s\gamma))$ and
$\epsilon_1$ in the no-scale $SU(5)\times U(1)$ SUGRA for the running top mass
$m_t=180\GeV$ 
(top row) and $m_t=170\GeV$ (bottom row). 
Points between the two horizontal lines 
or below the horizontal line as pointed by the arrow
are allowed by the $b\rightarrow s\gamma    $ constraint at the 95\% C.~L. while points between the two vertical lines are allowed by the
the $\epsilon_1$ constraint at the 90\% C.~L.
\end{itemize}

\end{document}